\RequirePackage{ifpdf}
\ifpdf 
\documentclass[pdftex]{sigma}
\else
\documentclass{sigma}
\fi

\def\fracpd#1#2{\frac{\partial #1}{\partial #2}}
\def\CL{{\cal L}}
\def\CX{{\cal X}}

\begin{document}

\renewcommand{\PaperNumber}{004}

\FirstPageHeading

\ShortArticleName{Symmetries of a System with Magnetic Terms}

\ArticleName{A System of $\boldsymbol{n=3}$ Coupled Oscillators with Magnetic Terms:
Symmetries and Integrals of Motion}

\Author{Manuel F. RA\~NADA}
\AuthorNameForHeading{M.F. Ra\~nada}

\Address{Departamento de F\'{\i}sica Te\'orica, Facultad de Ciencias,
Universidad de Zaragoza,\\ 50009 Zaragoza, Spain}
\Email{\href{mailto:mfran@posta.unizar.es}{mfran@posta.unizar.es}}

\ArticleDates{Received July 06, 2005, in final form September 16, 2005; Published online September 20, 2005}

\Abstract{The properties of a system of $n=3$ coupled oscillators
with linear terms in the velocities (magnetic terms) depending in
two parameters are studied. We proved the existence of a
bi-Hamiltonian structure arising from a non-symplectic symmetry,
as well the existence of master symmetries and additional
integrals of motion (weak superintegrability) for certain
particular values of the parameters. }

\Keywords{non-symplectic symmetries; bi-Hamiltonian structures;
master symmetries; \mbox{cubic} integrals}

\Classification{37J15; 70H06; 70H33}

\section{Introduction}

 The main objective of this paper is to present a study of a
system that, although arising from a quadratic Hamiltonian of simple form,
it is however endowed with several interesting properties related
with the theory of nonstandard symmetries and the existence of cubic
integrals of motion.

 Let us consider the Lagrangian of a system of $n=3$ coupled oscillators
with an additional coupling term linear in the velocities
(the so-called magnetic terms)
\begin{gather}
  L   =  \frac{1}{2} \big(v_1^2 + v_2^2 + v_3^2\big) - (V_1 + V_0), \nonumber\\
  V_0 = \frac{1}{2} k \big(q_{21}^2 + q_{32}^2 + q_{13}^2\big), \nonumber\\
  V_1 = b (q_{32} v_1 + q_{13} v_2 + q_{21} v_3) ,  \label{LV0V1}
\end{gather}
where $b$ and $k$ are arbitrary constants, $q_{ij}$ denote $q_i-q_j$,
and we assume that, for easiness of notation, the mass $m$ of the particle
is set equal to unity.
See Refs. \cite{DaS00} and \cite{McW00} for two recent papers dealing
with similar systems.

The starting point is to consider $L_0=T-V_0$, where $T$ denotes the
kinetic energy, as the basic original system, and the additional
term $V_1$ (linear in the velocities) as introducing a deformation.
We try to formulate the results in explicit dependence of $b$ in order
to study the changes of the dynamics when the parameter $b$
varies.

The Lagrangian $L$ has two exact Noether symmetries generated by the
vector fields
\begin{gather*}
  Y_2 = \fracpd{}{q_1} + \fracpd{}{q_2} + \fracpd{}{q_3}, \\
  Y_3 = q_{32} \fracpd{}{q_1} + q_{13} \fracpd{}{q_2}
+ q_{21} \fracpd{}{q_3},
\end{gather*}
with associated constants of the motion
\begin{gather*}
  I_2 = v_1 + v_2 + v_3,  \\
  I_3 = q_{32} v_1 + q_{13} v_2 + q_{21} v_3
          - b \big(q_{21}^2 + q_{32}^2 + q_{13}^2\big).
\end{gather*}
The first constant $I_1$ is the total energy given by
\begin{gather*}
 I_1 = \Delta(L) - L = \frac{1}{2} \big(v_1^2 + v_2^2 + v_3^2\big) +
     \frac{1}{2} k \big(q_{21}^2 + q_{32}^2 + q_{13}^2\big).
\end{gather*}
where $\Delta = v_i (\partial/\partial v_i)$ represents the Liouville vector field.

The Legendre transformation
\begin{gather*}
 p_1 = v_1 - b q_{32},\qquad
 p_2 = v_2 - b q_{13},\qquad
 p_3 = v_3 - b q_{21}
\end{gather*}
leads to the following Hamiltonian
\begin{gather}
  H =\frac{1}{2}\big[(p_1+bq_{32})^2 + (p_2+bq_{13})^2 + (p_3+bq_{21})^2\big]
+ \frac{1}{2}k\big(q_{21}^2 + q_{32}^2 + q_{13}^2\big)\nonumber\\
\phantom{H}{}= \frac{1}{2}\big(p_1^2 + p_2^2 + p_3^2\big) + b(L_1 + L_2 + L_3)
+ \frac{1}{2} (b^2+k)\big(q_{21}^2 + q_{32}^2 + q_{13}^2\big)  \label{HV0V1}
\end{gather}
and the above three constants become
\begin{gather*}
 J_1 = H,\qquad  J_2 = p_1 + p_2 + p_3 , \qquad
 J_3 = L_1 + L_2 + L_3,
\end{gather*}
where $L_i$, $i=1,2,3$, denote the components of the angular momentum
$L_i = q_j p_k - q_k p_j$.
This three integrals satisfy $\{J_r,J_s\} = 0$, $r,s=1,2,3$,
and make of $H$ an integrable system for all the values of $b$ and $k$.
Thus, we arrive to the following result

\begin{proposition}
If we consider the magnetic term $V_1$ as a continuous deformation
of the ori\-ginal system $L_0=T-V_0$ (the value of $b$ representing
the intensity of the deformation) then this deformation preserves
both the existence of Noether symmetries and the integrability of
the original undeformed system.
\end{proposition}

Notice that $H$ can also be written as $H = J_0 + bJ_3$ with
$J_0$ defined as follows
\begin{gather*}
 J_0 = \frac{1}{2}\big(p_1^2 + p_2^2 + p_3^2\big) + \frac{1}{2}
 (b^2+k)\big(q_{21}^2 + q_{32}^2 + q_{13}^2\big).
\end{gather*}

\section{Symmetries and integrals of motion}

In differential geometric terms, the dynamics of a time-independent
Hamiltonian system is determined by a vector field on the
$2n$--dimensional cotangent bundle $T^*Q$ of a $n$-dimensional manifold $Q$.
Cotangent bundles are manifolds endowed, in a natural or canonical way,
with a~symplectic structure $\omega_0$ that, in coordinates
$\{(q_j,p_j);\; j=1,2,\ldots,n\}$,
is given by
\begin{gather*}
  \omega_0 =  dq_j \wedge dp_j,\qquad
  \omega_0 = -d {\theta_0}, \qquad \theta_0 = p_j dq_j
\end{gather*}
(we write all the indices as subscripts and we use the summation
convention on the repeated index).
Given a differentiable function $F=F(q,p)$, the vector field $X_F$
defined as the solution of the equation
\begin{gather*}
 i(X_F) \omega_0 = dF
\end{gather*}
is called the Hamiltonian vector field of the function $F$.
There are two important properties:
\begin{enumerate}\itemsep=0pt
\item[(i)] The Hamiltonian vector field of a given function is well
defined without ambiguiti\-es.
This uniqueness is a consequence of the  symplectic character of
the two--form $\omega_0$.

\item[(ii)] Suppose that we are given a Hamiltonian $H=H(q,p)$.
Then the dynamics is given by the Hamiltonian vector field $\Gamma_H$
of the Hamiltonian function.
That is, $i(\Gamma_H)\omega_0 = dH$.
\end{enumerate}

\subsection{Bi-Hamiltonian structure}

At this point we recall that a (infinitesimal)  dynamical symmetry
of a Hamiltonian system $(T^*Q,\omega_0,H)$ is a vector field $Y$
such that it satisfies $[Y,\Gamma_H]=0$.
On the other hand it is known that, in some very particular cases,
the Hamiltonian systems can admit dynamical but non-symplectic symmetries
(for a classification of the symmetries in geometric terms see \cite{Cr83}
and~\cite{Pr83}).
In this case we have the following property.

\begin{proposition}
Suppose there is a vector field  $Y$ that is a dynamical symmetry of $\Gamma_H$
but does not preserve the symplectic two-form
\begin{gather*}
  \CL_{Y} \omega_0 = \omega_Y \ne 0.
\end{gather*}
Then (i) the dynamical vector field $\Gamma_H$ is bi-Hamiltonian,
and (ii) the function $Y(H)$ is the new Hamiltonian, and therefore it
is a constant of motion.
\end{proposition}

\begin{proof}
For a proof of this proposition see \cite{CaI83, CaMaRa02, Ra00},
and references therein.
A similar property is studied in \cite{Se04, Sm97} for the case of
Poisson manifolds.
\end{proof}

 Let us denote by $Y_3$ the Hamiltonian vector field of the function $J_3$
\begin{gather}
  i(Y_3)\omega_0 = d J_3 .  \label{DefY3}
\end{gather}
Then we have the following property:

\begin{proposition}
The vector field $Y_3$, defined by \eqref{DefY3} as the canonical infinitesimal
symmetry associated with $J_3$, can be written as a linear combination of two
dynamical but non-symplectic symmetries of $\Gamma_H$.
\end{proposition}

\begin{proof}
  We can write $Y_3$ as follows
\begin{gather*}
  Y_3 = Y_3^a - Y_3^b,\\
  Y_3^a  =  q_{3} \fracpd{}{q_1} + q_{1} \fracpd{}{q_2} + q_{2} \fracpd{}{q_3}
         +  p_{3} \fracpd{}{p_1} + p_{1} \fracpd{}{p_2} + p_{2} \fracpd{}{p_3} ,\\
  Y_3^b  =  q_{2} \fracpd{}{q_1} + q_{3} \fracpd{}{q_2} + q_{1} \fracpd{}{q_3}
         +  p_{2} \fracpd{}{p_1} + p_{3} \fracpd{}{p_2} + p_{1} \fracpd{}{p_3} .
\end{gather*}
Then the vector field $Y_3^a$ (or $Y_3^b$) is neither locally-Hamiltonian
with respect to $\omega_0$
\begin{gather*}
  \CL_{Y_3^a}\omega_0 =  \omega_a \ne 0   ,
\end{gather*}
nor a symmetry of the Hamiltonian function, $Y_3^a(H)\ne 0$.
Nevertheless it satisfies
\begin{gather*}
  [Y_3^a,\Gamma_H] = 0   .
\end{gather*}
Therefore $Y_3^a$ (or $Y_3^b$) is  a dynamical but non-symplectic symmetry
of the Hamiltonian sys\-tem.\end{proof}

{\samepage Thus the dynamical vector field $\Gamma_H$ is a bi-Hamiltonian system with
respect to $(\omega_0,\omega_a)$
\begin{gather}
  i(\Gamma_H)\omega_0 = d H  ,\qquad  i(\Gamma_H)\omega_a = d H_a ,
\end{gather}
with $\omega_a$ and $H_a = Y_a(H)$ given by
\begin{gather*}
 \omega_a  =  dq_1{\wedge}(dp_2+dp_3) + dq_2{\wedge}(dp_1+dp_3)
          + dq_3{\wedge}(dp_1+dp_3)   ,  \\
 H_a  =  \frac 12 J_2^2  -H     .
\end{gather*}
The function $H_a = Y_a(H)$ can be considered as a new Hamiltonian for $\Gamma_H$.}

The bi-Hamiltonian structure $(\Gamma_H,\omega_0,\omega_a)$ determines a $(1,1)$-tensor field
$R$ by the relation
\begin{gather*}
 \omega_a(X,Y)=\omega_0(RX,Y) , \qquad
 \forall\;  X,Y\in \CX(T^*Q) ,
\end{gather*}
or, equivalently, $R = \widehat{\omega}_0^{-1}{\circ}\widehat{\omega}_a$
[with $\widehat{\omega}(X) = i(X)\omega\in\wedge^1(T^*Q)$].
We have obtained
\begin{gather}
  R  =  \fracpd{}{q_1}{\otimes}(dq_2+dq_3) + \fracpd{}{q_2}{\otimes}(dq_1+dq_3)
     + \fracpd{}{q_3}{\otimes}(dq_1+dq_2)  \nonumber  \\
\phantom{R=}{} + \fracpd{}{p_1}{\otimes}(dp_2+dp_3) + \fracpd{}{p_2}{\otimes}(dp_1+dp_3)
     + \fracpd{}{p_3}{\otimes}(dp_1+dp_2) .
\end{gather}

 Starting with the basic Hamiltonian system
$(\omega_0,\Gamma_0=\Gamma_H,dH_0=dH)$, and iterating $R$, we can
construct a sequence of 2-forms $\omega_k$, vector fields
$\Gamma_k$, and exact 1-forms $dH_k$, $k=1,2,\ldots$, defined by
$\widehat{\omega}_k=\widehat{\omega}_0\circ R^k$,
$\Gamma_k=R^k(\Gamma_0)$, and $dH_k=R^k(dH_0)$, that leads to
sequence of bi-Hamiltonian vector fields. The two first steps
become
\begin{gather*}
i(\Gamma_0)\omega_1 = i(\Gamma_1)\omega_0 = dH_1 ,\qquad
i(\Gamma_0)\omega_2 = i(\Gamma_1)\omega_1 =
    i(\Gamma_2)\omega_0 = dH_2 ,
\end{gather*}
where $\omega_2$, $\Gamma_2$, and $H_2$, are given by
\begin{gather*}
  \omega_2 = 2 \omega_0 + \omega_1  , \qquad
  \Gamma_2 = 2 \Gamma_0 + \Gamma_1  , \qquad
       H_2 = 2 H_0 + H_1            .
\end{gather*}
So, in this particular case, the sequence
$(\omega_k,\Gamma_k,dH_k)$ closes over itself after two steps.

\subsection{Master symmetries}

The function $T=T(q,p)$ is said to be a generator of constants of
motion of degree $m$ if it is not preserved by the dynamics but it
generates an integral of motion by time derivation
\begin{gather*}
  \frac{d}{dt}T\ne0 ,\quad  \ldots,\quad \frac{d^{m-1}}{dt^{m-1}}T\ne 0,
\quad \frac{d^m}{dt^m}T= 0.
\end{gather*}
In differential geometric terms, a vector field $X$ on $T^*Q$ that satisfies
\begin{gather*}
  [\Gamma_H,X] \ne0,\qquad
  [\Gamma_H,{\widetilde X}] = 0  ,\qquad {\widetilde X} = [\Gamma_H,X],
\end{gather*}
is called a ``master symmetry" or a ``generator of symmetries"
of degree $m=1$ for $\Gamma_H$ \cite{Ca02, Da93, Fe93, Ra99}.
If a Hamiltonian $H$ admits two of such generators, $T_1$, $T_2$,
(master integrals) then the function $I_{12}$ defined by
\begin{gather*}
  I_{12} = I_2 T_1 - I_1 T_2   , \qquad
  I_r = \frac{d}{dt}T_r      , \qquad  r=1,2,
\end{gather*}
is a new constant of motion (independent of $I_1$ and $I_2$).
In geometric terms, if we denote by $X_r$, the Hamiltonian
vector fields of $T_r$, then ${\widetilde X}_r$ is Hamiltonian as well,
and we have
\begin{gather*}
  i(X_r)\omega_0  = dT_r     , \qquad
  i({\widetilde X}_r)\omega_0 = d{\widetilde T}_r    , \qquad
  {\widetilde T}_r = \Gamma_H(T_r)     , \qquad  r=1,2,
\end{gather*}
so that the vector field $X_{12}$ defined as
\begin{gather*}
 X_{12} = T_1 {\widetilde X}_2 - T_2 {\widetilde X}_1
        + {\widetilde T}_2 X_1 - {\widetilde T}_1 X_2  , \qquad
 i(X_{12})\omega_0 = dI_{12},
\end{gather*}
is a symmetry of the system $(T^*Q,\omega_0,H)$ with the function
$I_{12}$ as its associated constant of motion.

The Hamiltonian vector field $X_1$ of the function $T_1=q_1+q_2+q_3$ satisfies
\begin{gather*}
i(X_1)\omega_0  = dT_1  , \qquad  [\Gamma_H,X_1] = {\widetilde X}_1  ,\qquad
 {\widetilde X}_1 = \fracpd{}{q_1} + \fracpd{}{q_2} + \fracpd{}{q_3},
 \qquad  [\Gamma_H,{\widetilde X}_1]  = 0   ,
\end{gather*}
so $X_1$ is a ``generator of symmetries" of degree $m=1$ for
\eqref{HV0V1}. Similarly if we consider $T_2=q_1p_1+q_2p_2+q_3p_3$
we obtain
\begin{gather*}
  i(X_2)\omega_0 = dT_2  , \qquad  [\Gamma_H,X_2] = {\widetilde X}_2
, \qquad  [\Gamma_H,{\widetilde X}_2]\stackrel{*}{=}0  ,
\end{gather*}
where the asterisk means that the Lie bracket vanishes only if  $k+b^2=0$.
Hence in the particular case $k+b^2=0$, the system is superintegrable with
an additional integral of the form $I_{12}$ that becomes
\begin{gather}
  J_4 \equiv I_{12} = L_1(p_2-p_3) + L_2(p_3-p_1) + L_3(p_1-p_2) .
\end{gather}

\subsection{Third-order integrals}

 Linear integrals of motion be obtained from exact Noether symmetries
and quadratic integrals from Hamilton--Jacobi (Schr\"odinger)
separability but third-order integrals must be obtaining by other
alternative procedures; because of this, the number of known
integrable systems admitting cubic in the momenta constants is
very limited \cite{Gr04,GrW02,McS04,Sh04,Th84}. For $n=2$ we can point out
the Fokas--Lagerstrom and the Holt potentials for the Euclidean
plane \cite{FoL80,Ho82}, and three of the Drach potentials for the
pseudoeuclidean plane \cite{Ra97,RaS01}. For $n=3$ two very
special systems are well known: the three-particle Toda chain and
three-particle Calogero--Moser system which are both endowed with a
cubic constant of the form
\begin{gather}
  J  = \frac{1}{3}\big(p_1^3 + p_2^3 + p_3^3\big)  +
      {\rm terms\ of\ first\ order}     .    \label{J4TCM}
\end{gather}
At this point we recall the following property: polynomial
constants of the motion for mechanical systems (Riemmanian metric
with a potential) are either even or odd in the momenta
(for a~discussion of this result see, e.g., \cite{Hi87}).
Therefore if $J$ is a cubic integral for a Hamiltonian such as the
Calogero--Moser or the Toda systems then it must also contain
linear terms but neither quadratic nor independent ones.
Nevertheless this property is not true for the more general case
of a non-mechanical Lagrangian. In our case, due to the presence
of linear terms in the velocities (magnetic terms) when looking
for third-order integrals we must assume for the function $J_4$ a
polynomial expression containing also even powers.

 We start with a third-order function $J$ of a general form
\begin{gather*}
 J = \sum_{i+j+k=3} a_{ijk}p_1^i p_2^j p_3^k
   + \sum_{i+j+k=2} b_{ijk}p_1^i p_2^j p_3^k
   + \sum_{i} c_ip_i + d,
\end{gather*}
where $a_{ijk}$, $b_{ijk}$, $c_i$, and $d$ are functions of $q_1$, $q_2$, $q_3$.
Then the condition $\{H,J\}=0$ leads to a system of equations;
some of these equations restrict the expressions of the coefficients~$a_{ijk}$,
$i+j+k=3$, of the higher-order terms; the other equations couple
the derivatives of~$V_1$ and $V_0$ with the coefficients $a_{ijk}$,
$b_{ijk}$, $c_i$, and $d$ or with the derivatives of these coefficients.
Unfortunately, the general case is difficult to be studied but some
particular cases can be analyzed assuming simple particular forms for
the coefficients $a_{ijk}$.

  We have studied the existence of a cubic integral of such
a particular form for this three-particle system and we have
arrived (we omit the details) at the existence of the following
function
\begin{gather}
  J_4   =   \frac{1}{3} \big(p_1^3 + p_2^3 + p_3^3\big)
      -  b \big(q_{32} p_1^2 + q_{13} p_2^2 + q_{21} p_3^2\big) \nonumber\\
\phantom{J_4   =}{} +   b^2 (z_1 p_1 + z_2 p_2 + z_3 p_3)
     + 9 b^3 \big(q_{21}^3 + q_{13}^3 + q_{32}^3\big) ,     \label{J4}
\end{gather}
where $z_1$, $z_2$, and $z_3$, are given by
\begin{gather*}
  z_1 = 4 \big(q_{21}^2 + q_{13}^2\big) + q_{32}^2  ,\qquad
  z_2 = 4 \big(q_{32}^2 + q_{21}^2\big) + q_{13}^2  ,\qquad
  z_3 = 4 \big(q_{13}^2 + q_{32}^2\big) + q_{21}^2  .
\end{gather*}
The important point is that the function $J_4$ is a constant of
motion, not for all the values of the parameters $(b,k)$, but only
in the particular case of $k$ and $b$ satisfying $k=8b^2$. We
have verified that in this case the four functions $H$, $J_2$,
$J_3$, $J_4$ are independent and that the system ($J_2$, $H$,~$J_4$) is involutive (nevertheless  $\{J_3,J_4\} \ne 0$).
Thus, when $k$ and $b$ are related by $k=8b^2$ then the system
becomes superintegrable with an additional third-order integral.

We finally note that $J_2$ is linear, $H$ quadratic, and $J_4$ cubic,
and that they have expressions of the form
\begin{gather}
 J_2 =  p_1 + p_2 + p_3,    \qquad
 H   =  p_1^2 + p_2^2 + p_3^2  + \cdots, \qquad
 J_4  =  p_1^3 + p_2^3 + p_3^3  + \cdots
\end{gather}
that closely resemble the expressions of the three first constants of
Toda or the Calogero--Moser systems.

\section{Final comments}

The Lagrangian system \eqref{LV0V1} (Hamiltonian \eqref{HV0V1}) is
a system endowed with a bi-Hamiltonian structure, non-symplectic
symmetries and master symmetries. Moreover for a particular value
of the parameter $b$, it becomes superintegrable with a cubic
integral.

 The non-symplectic symmetry $Y_3^a$ is $b$-independent and hence
the bi-Hamiltonian structure $(\Gamma_H,\omega_0,\omega_a)$ is the
one endowed by the the original oscillators system $L_0=T-V_0$
that is preserved under the $b$-deformation introduced by the
magnetic term $V_1$. Concerning the master symmetries, the first
one is always present but the second only exists if
$k+b^2=0$. So in this particular case the system becomes (weakly)
superintegrable. This particular case is very peculiar because the
Hamiltonian \eqref{HV0V1} becomes a system of $n=3$ particles with
velocity-linear couplings but without the potential term.

Finally the particular $k=8b^2$ case is very interesting. First
because, as stated above, the number of known integrable systems
with third-order integrals is very small. Second because the
existence of this property for a very particular case of the
parameters resembles situations characterizing other known systems
(e.g., the H\'enon--Heiles system is only integrable for some very
particular values of a parameter). Finally because the expression
obtained for $J_4$ in \eqref{J4} shows similarity with the
third-order integral \eqref{J4TCM} of the Toda or the
Calogero--Moser systems. This leads to the necessity of the study
of the more general $n>3$ case, that poses the question of the
nature of the interaction (in the Toda chain the interactions are
between neighbours but in the Calogero--Moser case they are between
every two particles). On the other hand this also leads to the
search of a Lax pair generating the three functions ($J_2$, $H$,
$J_4$), as traces of the powers of an appropriate matrix.
Nevertheless, notice that this Lax structure (if it exists) must
correspond  to the $L$ system with the additional condition $k=8b^2$;
so it cannot be obtained as a continuous deformation of a
previous Lax pair for the $L_0$ system.
We think that these are open questions that must be
investigated.

\subsection*{Acknowledgements}

Support of projects  BFM-2003-02532 and FPA-2003-02948
is acknowledged.

\LastPageEnding

\end{document}